\documentclass[preprint,5p,times,twocolumn]{elsarticle}

\usepackage{graphicx}
\usepackage{epstopdf}
\usepackage{color}
\usepackage{bm}
\usepackage{amssymb}
\usepackage{amsfonts}
\usepackage{amsmath}
\usepackage{float}
\usepackage{siunitx}

\journal{Carbon}

\begin{document}
\begin{frontmatter}

  \title{Charge-induced electrochemical actuation of armchair carbon
    nanotube bundles}

  \author[a]{Nguyen T. Hung\corref{cor1}} \cortext[cor1]{Corresponding
    author. Tel.:+81 22 795 7754; Fax: +81 22 795 6447.}
  \ead{nguyen@flex.phys.tohoku.ac.jp}

  \author[a]{Ahmad R. T. Nugraha}

  \author[a]{Riichiro Saito}

\address[a]{Department of Physics, Tohoku University,
    Sendai 980-8578, Japan}

\begin{abstract}
  The effects of charge doping on the structural deformation and on
  the electronic structure of armchair single wall carbon nanotube
  (SWNT) bundles are investigated through first-principles
  calculations.  In particular, we select a $(6, 6)$ SWNT as an
  example and we calculate a mechanical deformation in the SWNT
  bundles as a function of gate voltage, which could serve as a basis
  of the electromechanical actuators in an artificial muscle.  We find
  that the magnitudes of the actuation responses such as strain and
  stress of the $(6, 6)$ SWNT bundle in the case of hole doping are
  substantially larger than those of electron doping.  The $(6, 6)$
  SWNT bundle also exhibits a low-symmetry and opens an energy band
  gap of about 0.41 eV around the charge neutral condition, which
  allows a semiconductor-to-metal transition in the electron-doping
  regime when the relative shift of the Fermi energy goes up to 0.60
  eV, above which the Young modulus increases.
\end{abstract}


\end{frontmatter}

\section{Introduction}
An artificial muscle is a novel actuator made of a variety of polymers
or composites that change their shape when excited electrically.  A
composite of single wall carbon nanotubes (SWNTs) is known to be one
of the strongest and stiffest
materials~\cite{yu2000strength,yu2000tensile,hung2016intrinsic}, and
they can reversibly contract and expand in volume under an applied
voltage, similar to the natural muscle.  Almost two decades ago,
Baughman et al.~\cite{baughman1999carbon,baughman2002carbon} had built
an ultra-strong artificial muscle from bundles (or yarns) of SWNTs
with electrochemical charging, in which the artificial muscle could
generate force per unit area of about $26$ MPa, which is almost $100$
times larger than that of natural muscle ($\approx 0.35$
MPa)~\cite{madden2004artificial}.  Therefore, the charge-induced
electromechanical actuators of the SWNT bundles are not only
fascinating from the scientific point of view but are also promising
for technological applications.  Gartstein et
al.~\cite{gartstein2002charge} analytically showed that the magnitude
of the actuation response of semiconducting SWNTs could be
substantially larger than that of graphite.  On the other hand, Yin et
al.~\cite{yin2011fermi} showed by first-principles calculation that
carbon-carbon bond lengths of metallic SWNTs are changed by charge
doping level, which lead to the modification of their electronic
properties.  While the artificial muscles in experiments were made of
SWNT \emph{bundles} (whose electronic structures show lower symmetry
than those of the individual SWNTs), the theoretical descriptions for
both the electronic structure and deformation under electromechanical
charging so far are only available for \emph{individual}
SWNTs~\cite{gartstein2002charge,yin2011fermi,yin2010fermi,li2007theoretical}.

A nanotube bundle consists of tens or several hundred individual SWNTs
arranged in a hexagonal lattice of tubes and bound by the van der
Waals force.  A bundle containing many nanotubes with only one
chirality is especially interesting because of its expected uniform
response as an artificial muscle.  The chirality of a SWNT is defined
by a set of integers $(n,m)$ which specifies the geometrical structure
of a SWNT and hence its physical
properties~\cite{saito1992electronic}.  With recent advances in the
SWNT fabrication technique, armchair $(n,n)$ and zigzag $(n,0)$ SWNT
bundles can be selectively
synthesized~\cite{blum2011selective,haroz2010enrichment}.  Moreover,
separation and purification of SWNTs into single chirality level are
now possible~\cite{liu2011large} and the alignment of SWNTs has also
been realized~\cite{shaver2008alignment}.  Thus the computational
design of artificial muscle will be important for obtaining the
optimum artificial muscle.  Note that although the formation of
individual SWNTs into a bundle involves weak van der Waals
interactions, their electronic properties can change significantly.
For example, by a first principles calculation, Delaney~\emph{et al.}
showed that a lowering symmetry of the armchair SWNT bundles caused by
interactions between tubes induces a small energy band gap although
the isolated armchair SWNTs are metallic
tubes~\cite{delaney1998broken}, which might affect the response of
artificial muscle.

Among different SWNT structures, the armchair SWNTs are interesting to
be studied for artificial muscle applications because of the two
following reasons.  First, an appropriate selection of the chirality
of the $(6, 6)$ armchair SWNTs could match the hexagonal lattice of
the bundle which saves the computational time.  Second, the armchair
SWNTs are basically known to be metallic and the effect of the bundle
formation on the SWNT electronic structure is more pronounced for
metallic SWNTs, compared with semiconducting ones.  Combining these
two reasons, we could suggest the optimum structure for the artificial
muscles.  Furthermore, due to the weak tube-tube interactions, it is
important to study the effects of electrochemical doping or gate
voltage on the symmetry and on the electronic properties of the
armchair SWNT bundles.  However, many researchers often assumed in the
their calculation that the energy bands do not change by doping level
and that only the Fermi energy is shifted by doping, which is known as
the rigid band model~\cite{lankhorst96-rigidband,marshak84-rigidband}.
Even for isolated and individual SWNTs, it is expected that the rigid
band model is not suitable to consider the heavy doping of the
metallic SWNTs~\cite{yin2011fermi}, in which the lattice constant is
modified.

In this paper, by considering the changes in the energy bands, we
discuss the electromechanical properties of the armchair SWNT bundles
as a function of charge doping, for both electron and hole doping
cases.  By first-principles calculation, we find two important
physical phenomena in the doped armchair SWNT bundles, one is the
semiconductor-metal transition in the armchair SWNT bundles as a
result of heavy electron doping, and another one is a large change of
strain by hole doping, which is essential in the artificial muscle
applications.  We expect that the armchair SWNT bundles can be
strained by the doping due to the extremely high Young's modulus of
individual SWNTs formed into the bundles~\cite{baughman1999carbon}.
For individual SWNTs and graphite, previous theoretical studies showed
that an axial strain of about 1\% might be achieved for both heavy
electron and hole doping
cases~\cite{mirfakhrai2008electromechanical,sun2002dimensional}.
Meanwhile, some previous experiments with the SWNT bundles showed that
the largest strain is $-1$\% when the negative potential is applied
(hole-doped) and that the strain is very small ($\sim 0.02\%$) when
the positive potential is applied
(electron-doped)~\cite{mirfakhrai2007electrochemical,foroughi2011torsional},
whose origin is not clear yet to our knowledge.  In this work, we
reproduce such an asymmetric behavior of the strain through the
calculations of the strain, Young's modulus, and stress of the
armchair SWNT bundles as a function of Fermi energy.  We will show
that the lengths of C-C bonds parallel and perpendicular to the bundle
axis change differently for the hole doping and electron doping.

\section{Calculation methods}
To calculate the structural deformation variables (strain, Young's
modulus, and stress) as a function of the Fermi energy, we define the
relative shift of the Fermi energy, $\Delta E_\mathrm F$, for the
charge (electron and hole) doping, which can be expressed by
\begin{equation}
\label{eq:1}
\Delta E_\mathrm F=E_{\mathrm F}^{\text{neutral}}-E_{\mathrm F}^{\text{doped}},
\end{equation}
where $E_{\mathrm F}^{\text{neutral}}$ and
$E_{\mathrm F}^{\text{doped}}$ are the Fermi energy for a neutral and
a charge doping, respectively.  The Fermi energy is defined by the
center of the energy gap for the semiconducting SWNT bundle case.  In
the case of the metallic SWNT bundle, in which the conduction and
valence energy bands touch each other at the Dirac point, the Fermi
energy is defined as the highest energy for occupied electrons.  Note
that equation~\eqref{eq:1} was already given in some earlier
studies~\cite{yin2011fermi,yin2010fermi}, discussing the Fermi level
dependence of optical transition energy and electronic properties in
the SWNTs.

\begin{figure}[t]
  \centering
  \includegraphics[clip,width=8.5cm]{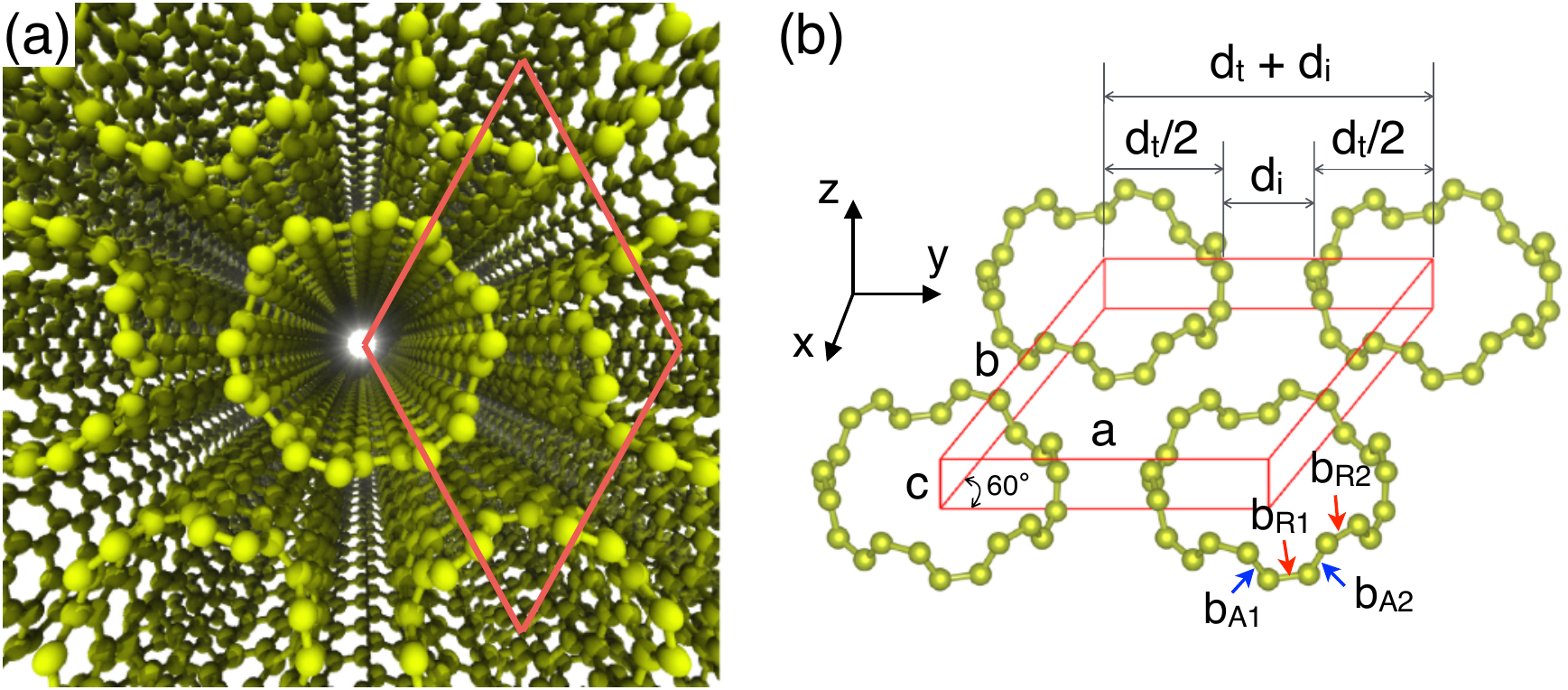}
  \caption{\label{fig:model} (a) Perspective view of the $(6, 6)$
    armchair SWNT bundle in three-dimensional space. (b) Hexagonal
    unit cell of the $(6, 6)$ armchair SWNT bundle showing the lattice
    vector $c$ along the tube axis, the diameter $d_t$, the intertube
    distance $d_i$, the C-C bonds perpendicular to the tube axis
    ($b_{R1}$ and $b_{R2}$), and those parallel to the tube axis
    ($b_{A1}$ and $b_{A2}$).  Note that the box in (b) is a rhombus,
    thus the $x$-axis is not parallel to any edges of the box.}
\end{figure}

In this work, we choose the chirality $(n,m) = (6, 6)$ for a model
calculation, which has the highest symmetry in the bundle, similar
with that in the isolated SWNT.  In Fig.~\ref{fig:model}(a) and (b),
we show the perspective view of a bundle of $(6, 6)$ armchair SWNTs in
three-dimensional space and its hexagonal unit cell including 24
carbon atoms, respectively.  The periodic boundary conditions are
applied for three dimensions.  The simulation hexagonal unit cell
dimensions are $a=b=d_t+d_i$ and $c$, where $c$ (in \AA) is the length
of the unit vector along the tube axis, $d_t$ (in \AA) is the diameter
of the tube, and $d_i$ (in \AA) is the intertube distance as shown in
Fig.~\ref{fig:model}(b).

We calculate the electronic properties of the $(6, 6)$ armchair SWNT
bundles from first-principles by using \texttt{Quantum
  ESPRESSO}~\cite{giannozzi2009quantum}, which is a full density
functional theory (DFT) simulation package using a plane-wave basis
set~\cite{hohenberg1964inhomogeneous,kohn1965self}.  The
Rabe-Rappe-Kaxiras-Joannopoulos ultrasoft pseudopotential with an
energy cutoff of 60 Ry is chosen for the expansion of the plane
waves~\cite{rappe1990optimized}.  The exchange-correlation energy is
evaluated by the general-gradient approximation using the
Perdew-Burke-Ernzerhof (PBE)
function~\cite{pseudo,perdew1996generalized}, which is appropriate to
model the interactions between the tubes in the bundle.  We use a
second version of the nonlocal van der Waals functional
(vdW-DF2)~\cite{thonhauser2007van}.  Using vdW-DF2 is necessary to
correctly capture the van der Waals interaction between the tubes in
the bundle for obtaining the forces and the lattice
parameters~\cite{dumlich2011nanotube}, which are important parameters
for calculating the mechanical properties and electronic structure.
In our simulation, the $8 \times 8 \times 24$ \textbf{k}-point grids
in the Brillouin-zone are used following the Monkhorst-Pack scheme,
where \textbf{k} is the electron wave
vector~\cite{monkhorst1976special}.

To obtain the SWNT bundle structure, the atomic positions and cell
vectors are fully relaxed by using the
Broyden-Fretcher-Goldfarb-Shanno minimization
method~\cite{broyden1970convergence,fletcher1970new,goldfarb1970family,shanno1970conditioning}.
This model is considered to be optimized when all the Hellmann-Feynman
forces and all components of the stress are less than
$5.0 \times 10^{-4}$ Ry/a.u. and $5.0 \times 10^{-2}$ GPa,
respectively, which are adequate for the present purpose.  In
Fig.~\ref{fig:model}(b), we present scaled bond structures for the (6,
6) SWNTs in the bundle without the charge doping, where
$b_{R1}=b_{R2}=1.427$ \AA\ and $b_{A1}=b_{A2}=1.428$ \AA\ represent
the C-C bonds perpendicular to and \ang{30} from the tube axis,
respectively.  The tube orientation in the present model corresponds
to the AB stacking in graphite~\cite{okada2001pressure}.  It should be
noted that in the case of $(6, 6)$ armchair SWNT bundle, the AB
stacking configuration is more stable than the AA stacking
configuration~\cite{dumlich2011nanotube,okada2001pressure}.

For the energy band calculations, we used $100~\textbf{k}$ points
along the $z$-direction.  To discuss the electromechanical actuation
of the $(6, 6)$ armchair SWNT bundle with the charge doping level
($\Delta Q$), ranging from $-0.9e$ to $+2.0e$ per unit cell, the
electron (hole) doping is simulated by adding (removing) electrons to
the SWNT bundle with the same amount of uniform positive (negative)
charge in the background so as to keep the charge neutrality.

\section{Results and discussion}

\subsection{Electronic structure of SWNT bundle}

\begin{figure}[t]
  \centering
  \includegraphics[clip,width=8cm]{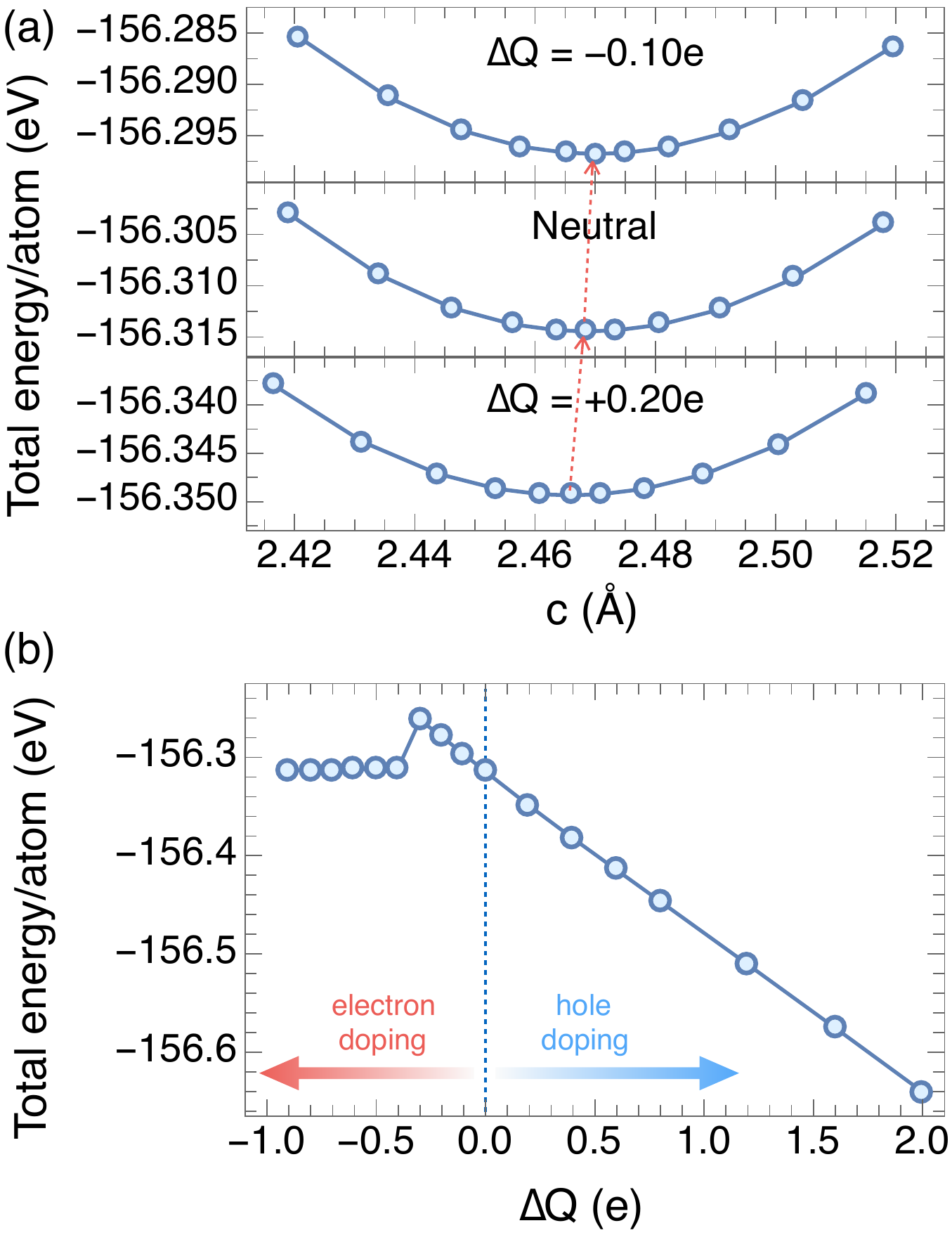}
  \caption{\label{fig:energy} (a) Total energy per atom of the
    $(6, 6)$ armchair SWNT bundle unit cell plotted as a function of
    the lattice constant $c$. (b) Minimum total energy per atom of the
    $(6, 6)$ armchair SWNT bundle as a function of the charge
    (electron and hole) doping, with $\Delta Q$ ranging from $-0.9e$
    to $+2.0e$ per unit cell.}
\end{figure}

In Fig.~\ref{fig:energy}(a), we show the total energy per atom of the
$(6, 6)$ armchair SWNT bundle as a function of the lattice vector $c$
ranging from 2.42 to 2.52 \AA. As shown in Fig.~\ref{fig:energy}(a),
the local minimum total energy is observed at $c=2.469$ \AA\ in the
neutral case which corresponds to the diameter $d_t=8.226$ \AA\ and
the intertube distance $d_i=3.258$~\AA\, which is consistent with the
previous reports~\cite{dumlich2011nanotube,okada2001pressure}. The
scanning of the potential energy surfaces as a function of lattice
constant are calculated for each charge (electron and hole) doping
$\Delta Q$.  In Fig.~\ref{fig:energy}(b), we show the minimum total
energy per atom of the $(6, 6)$ armchair SWNT bundle as a function of
$\Delta Q$ ranging from $-0.9e$ to $+2.0e$ per unit cell.  Since we
have 24 carbon atoms in the unit cell, $-0.1e$ ($+0.1e$) for electron
(hole) doping corresponds to the added (removed) 0.004167 electron per
carbon atom.  The total energy monotonically decreases with increasing
the hole doping.  A similar trend with sharper slope is also found for
the electron doping from $0.0e$ to $-0.3e$.  However, the total energy
becomes constant with further increase in the electron doping after a
finite jump of total energy at $\Delta Q=-0.4e$.  This suggests that
the atomic structure of the bundle may be broken into individual
SWNTs, which will be discussed later through intertube distance $d_i$
calculations for heavy electron doping case.

\begin{figure}[t]
  \centering
  \includegraphics[clip,width=8cm]{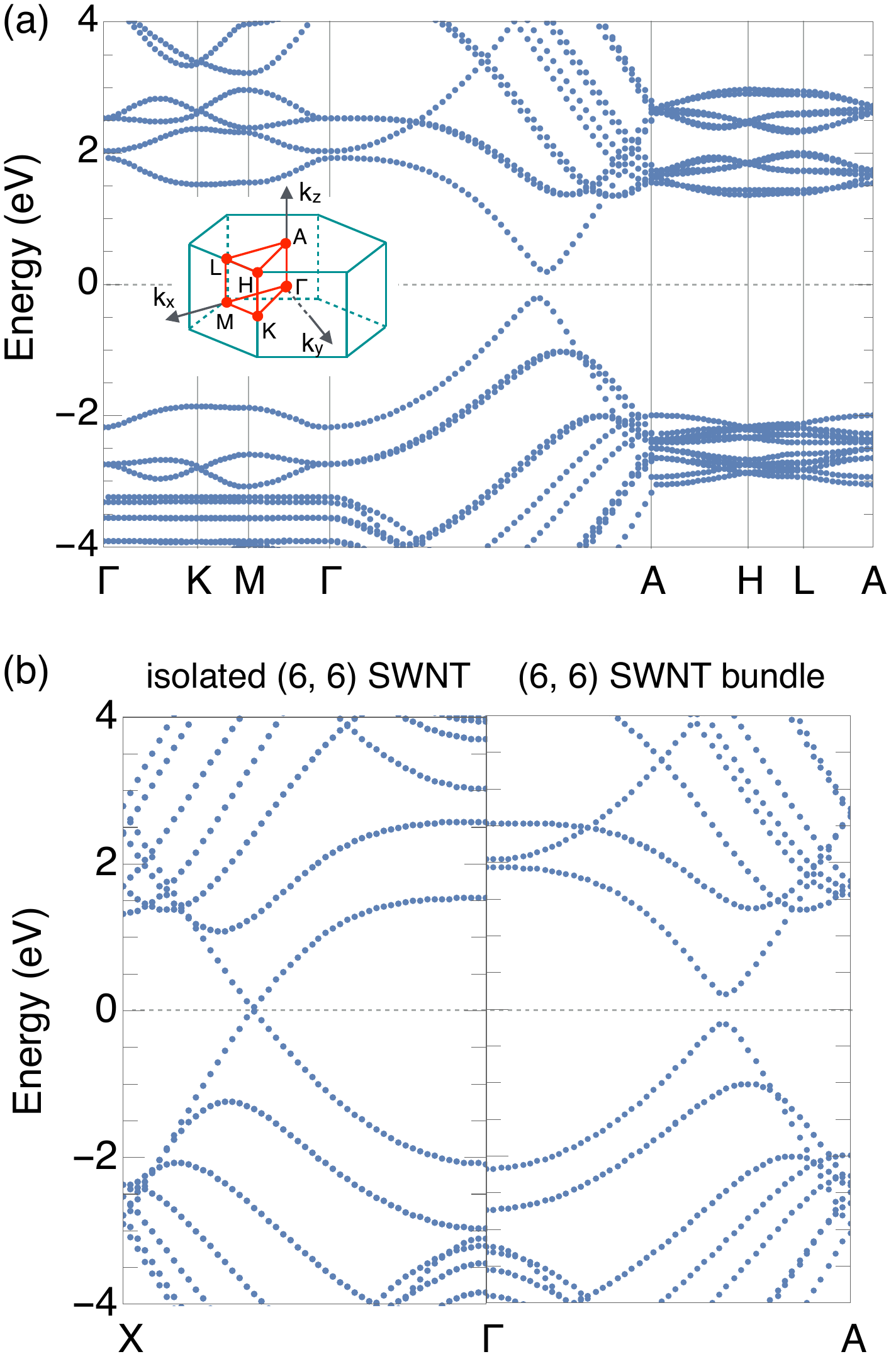}
  \caption{\label{fig:band}(a) Energy band structure along
    high-symmetry directions of the neutral $(6, 6)$ armchair SWNT
    bundle.  Inset shows the high-symmetry points and lines in the
    corresponding Brillouin zone.  (b) Energy band structure along
    $\Gamma$--$X$ and $\Gamma$--$A$ directions of the isolated
    $(6, 6)$ SWNT and the $(6, 6)$ SWNT bundle, respectively.}
\end{figure}

\begin{figure*}[ht]
  \centering
  \includegraphics[clip,width=13.5cm]{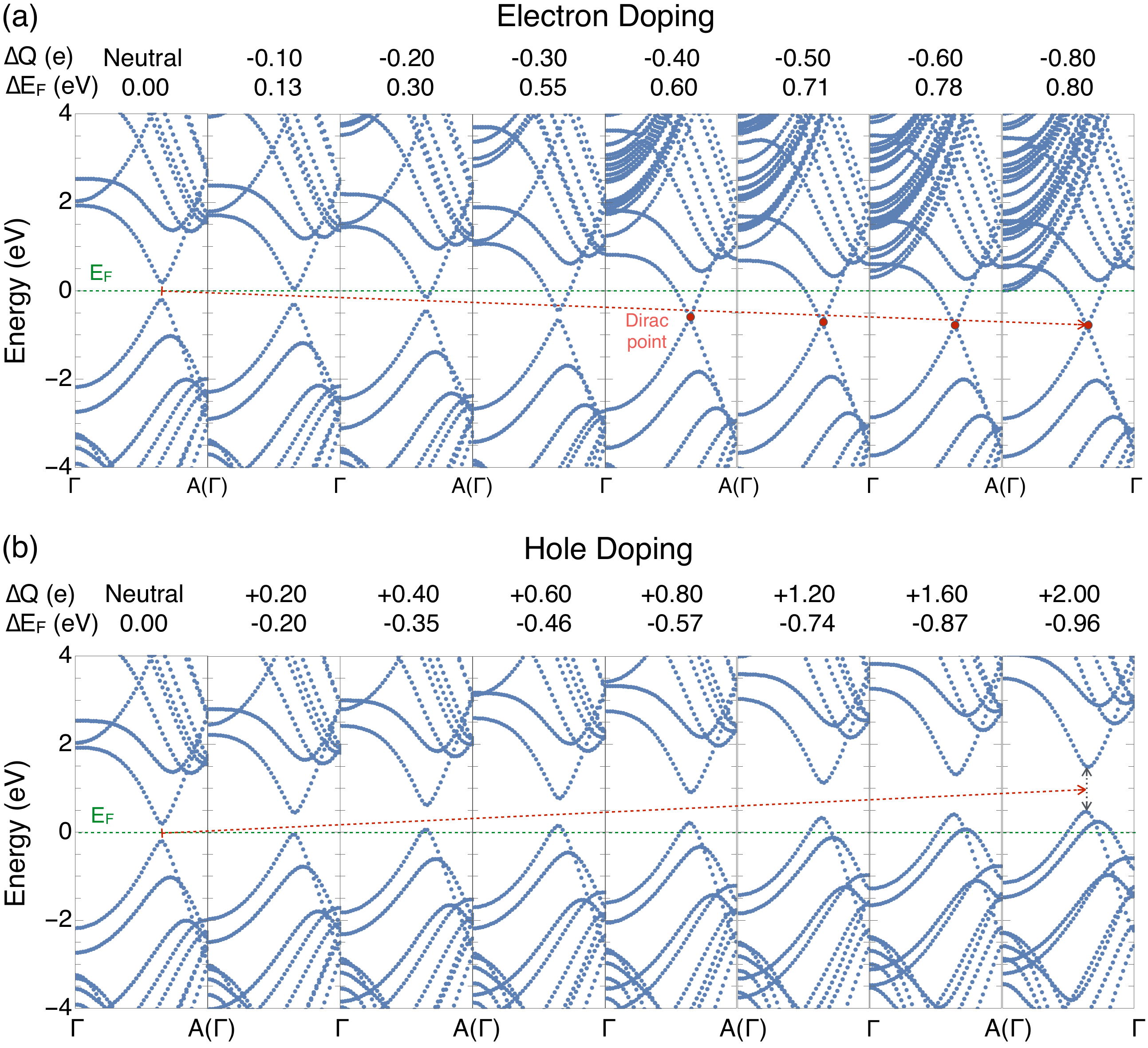}
  \caption{\label{fig:band6-6}Energy band structures of the $(6, 6)$
    armchair SWNT bundle with different (a) electron doping and (b)
    hole doping levels including those of the neutral $(6, 6)$
    armchair SWNT bundle for comparison.  The Fermi energy (straight
    dashed line) is set to zero for all plots.  Dashed arrows are a
    guide for eyes to see the evolution of the Dirac point.}
\end{figure*}

Before discussing the optimized electronic structure of the $(6, 6)$
armchair SWNT bundle, it is worth mentioning the symmetry properties
of individual armchair $(n,n)$ SWNTs.  The symmetry of an $(n,n)$ SWNT
is expressed by the direct product of the groups $D_n \otimes C_i$,
where $D_n$ consists of a vertical $n$-fold rotational axis $C_n$
(parallel to the nanotube axis) and $n$ horizontal 2-fold axes $C_2$
(perpendicular to the nanotube axis), and $C_i$ consists of the
identity $E$ and inversion $\sigma_i$
operators~\cite{saito1998physical}.  Since the symmetry operations
$D_n \otimes C_i$ are described by the symmetry group $D_{nh}$ or
$D_{nd}$ for even or odd $n$, respectively, an isolated individual
$(6, 6)$ armchair SWNT with $n=6$ has the $D_{6h}$ symmetry.  The
structure of the bundle has a comparably high-symmetry if the
individual nanotube and the bundle share all symmetry operations.
Therefore, the $(6, 6)$ armchair SWNT bundle can have either the
high-symmetry $D_{6h}$ or the low-symmetry $C_{6h}$ (loss of mirror
planes) corresponding to the AA stacking or AB stacking in graphene,
respectively~\cite{delaney1998broken,dumlich2011nanotube}.  Since the
AB stacking configuration is more stable than the AA stacking
configuration~\cite{dumlich2011nanotube,okada2001pressure}, we
hereafter consider only the structure of the $(6, 6)$ armchair SWNT
bundle with the AB stacking configuration with $C_{6h}$ symmetry.

Figure~\ref{fig:band}(a) displays the energy band structure along the
high-symmetry directions of the neutral $(6, 6)$ armchair SWNT bundle.
The inset shows the symmetry points and lines in the corresponding
Brillouin zone of the bundle.  The lowering of symmetry from $D_{6h}$
to $C_{6h}$ leads to a band-gap opening in the bundle of armchair
tubes~\cite{dumlich2011nanotube,delaney1998broken,reich2002electronic}.
As shown in Fig.~\ref{fig:band}(a), an energy band gap
$E_g\approx0.41$ eV opens up along the $\Gamma-A$ direction as a
result of the low-symmetry $C_{6h}$ in the $(6, 6)$ armchair SWNT
bundle with the AB stacking configuration.  This result reproduces a
previous theoretical report of Ref.~\cite{dumlich2011nanotube}. In
Fig.~\ref{fig:band}(b), we compare the energy band structure an
isolated $(6, 6)$ armchair SWNT with that of a neutral $(6, 6)$
armchair SWNT bundle considered in this work.  Since the isolated
$(6, 6)$ armchair SWNT has the high-symmetry $D_{6h}$, it shows a
metallic behavior without the energy band gap.  We note that the
metal-like behavior is expected for the high-symmetry with AA-stacked
configuration of the
bundle~\cite{dumlich2011nanotube,reich2002electronic}.  As we can see
from Fig.~\ref{fig:band}(b), the valence and the conduction bands of
the neutral $(6, 6)$ armchair SWNT bundle are shifted upward and are
more asymmetric compared with those of the isolated $(6, 6)$ armchair
SWNT. The asymmetric distribution of the conduction and the valence
bands originates from not only the effect of the overlap matrix
element and the curvature-induced $\sigma-\pi$
hybridization~\cite{ajayan1999nanotubes}, but also from the effect of
the van der Waals interaction between the tubes in the
bundle~\cite{dumlich2011nanotube}.

To study the variation of the electronic properties of the $(6, 6)$
armchair SWNT bundle with $\Delta Q$, we have calculated the energy
band structures of the $(6, 6)$ armchair SWNT bundle for all electron
(hole) doping levels considered in the present work.
Figures~\ref{fig:band6-6}(a) and (b) show the energy band structures
of the $(6, 6)$ armchair SWNT bundle under the electron and hole
doping, respectively. As shown in Fig.~\ref{fig:band6-6}(a), for the
electron doping, the energy bands shift downward compared with the
energy bands in the neutral case.  The downward-shift gives
$\Delta E_\mathrm F >0$.  Since the downward-shift of the conduction
bands is larger than that of the valence bands with decreasing
$\Delta Q$, the energy gap $E_g$ decreases and eventually becomes a
Dirac energy point at $\Delta Q = -0.40e$.  Thus, the $(6, 6)$
armchair SWNT bundle becomes metallic at the electron doping at at
$\Delta Q = -0.40e$.  For further electron doping, the Dirac point
slightly moves downward to a lower energy position.  In the case of
hole doping, the energy bands of the $(6, 6)$ armchair SWNT bundle
shift upward by increasing $\Delta Q$, as shown in
Fig.~\ref{fig:band6-6}(b), resulting in $\Delta E_\mathrm F <0$. The
upward-shifts of the valence bands are smaller than those of the
conduction bands with increasing $\Delta Q$, which makes an increase
of $E_g$ for the hole doping.  For heavy electron (hole) doping, the
second conduction (valence) subband touches the Fermi energy which
might change the transport properties of the SWNT bundle. We note
that the asymmetric distribution of the conduction and the valence
bands relative to the energy gap increases for both electron and hole
doping cases. 

As mentioned in the introduction, the rigid band model was commonly
considered to investigate the effect of charge doping on the
electronic structure of individual SWNTs or SWNT
bundles~\cite{okada2000nearly,ayala2010doping,hung2015diameter}. In
this model, the effective masses and band gaps are fixed, while the
Fermi levels are shifted by charge doping.  The applicability of the
rigid band model was justified, for instance, in alkali- or
nitrogen-doped SWNTs~\cite{ayala2010doping} and in the calculation of
thermoelectric power of the semiconducting
SWNTs~\cite{hung2015diameter}.  However, in the case of heavy doping
to obtain a high actuation, in which the Fermi energy can be changed
up to $1$ eV, the rigid band model is no longer suitable because the
energy structure changes significantly.  Therefore, in this work, we
did not use the rigid band model, but utilized the altered band
structures as they were obtained from the first-principles
calculations.

\begin{figure}[t]
  \centering
  \includegraphics[clip,width=8cm]{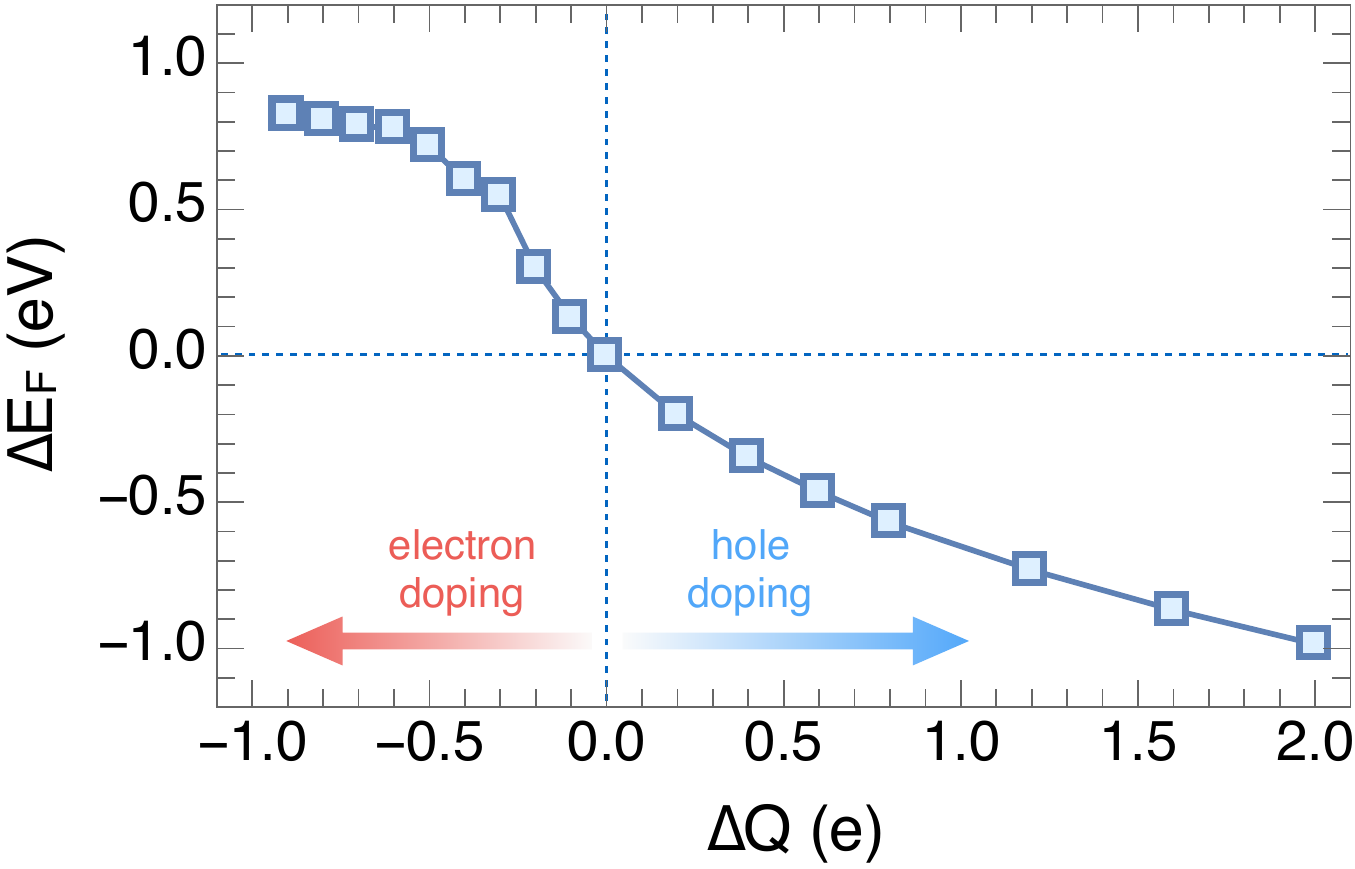}
  \caption{\label{fig:fermi-level}$\Delta E_{\mathrm F}$ as a function
    of $\Delta Q$ for charge (electron and hole) doping.}
\end{figure}

\begin{figure}[t]
  \centering
  \includegraphics[clip,width=8cm]{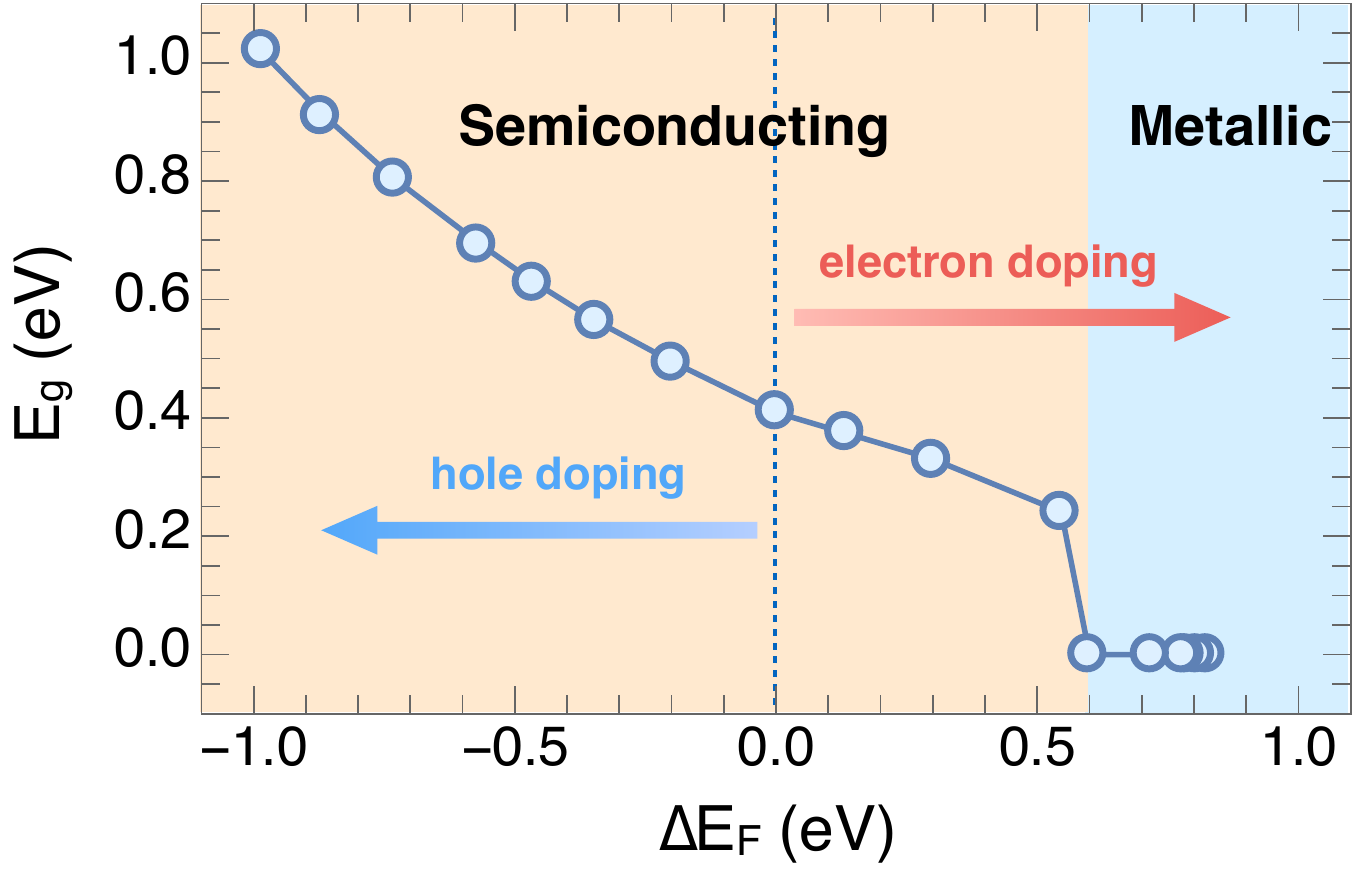}
  \caption{\label{fig:band-gap}Energy band gaps of the $(6, 6)$
    armchair SWNT bundle as a function of $\Delta E_{\mathrm F}$.  The
    apricot and light blue shadings indicate the regime in which the
    $(6, 6)$ armchair SWNT bundle has semiconducting and metallic
    states, respectively.}
\end{figure}

In Fig.~\ref{fig:fermi-level}, we show $\Delta E_\mathrm F$ as a
function of $\Delta Q$ based on Eq.~\eqref{eq:1}.  In the neutral
case, $\Delta E_\mathrm F = 0$ from the definition in
Eq.~\eqref{eq:1}, while for the electron (hole) doping,
$\Delta E_\mathrm F$ monotonically decreases with increasing
$\Delta Q$.  As we can see from Fig.~\ref{fig:fermi-level}, the change
of $\Delta E_\mathrm F$ with electron doping is larger than that with
hole doping because of an asymmetry of energy bands (see in
Fig.~\ref{fig:band6-6}).  For the heavy electron (hole) doping, the
higher energy valence bands around the $\Gamma$ point show smaller
downward (upward)-shift, resulting in a smaller increase (decrease) of
$\Delta E_\mathrm F$.  In the electron (hole) doping, $\Delta Q$ of
$-1.0e$ ($+2.0e$) per unit cell corresponds to the addition of 0.04167
(removal of 0.08334) electron per carbon atom, which is appropriate
for our calculation because the limit of the experimentally accessible
charge is from about $-0.3$ to $+0.1$ electron per carbon atom for
graphite~\cite{sun2002dimensional}.  In Fig.~\ref{fig:band-gap}, we
plot the energy gap $E_g$ of the $(6, 6)$ armchair SWNT bundle as a
function of $\Delta E_\mathrm F$.  In the neutral case, the $(6, 6)$
armchair SWNT bundle is semiconducting with $E_g=0.41$ eV. For the
electron (hole) doping, $E_g$ decrease (increase) with increasing
(decreasing) $\Delta E_\mathrm F$.  This result suggests that the
energy band gap of the $(6, 6)$ armchair SWNT bundle is tunable by
changing the Fermi energy, which can be achieved experimentally by
electrochemical doping or by applying a gate voltage.  As shown in
Fig.~\ref{fig:band-gap}, a semiconductor-metal transition occurs in
the $(6, 6)$ SWNT bundle when $\Delta E_\mathrm F$ is adjusted up to
0.60 eV. The reason for this transition will be discussed below in
terms of the C-C bond length, which is also related with the symmetry
properties of the $(6, 6)$ armchair SWNT bundle as a function of
$\Delta E_\mathrm F$.

\subsection{Mechanical deformation}

\begin{figure}[t]
  \centering
  \includegraphics[clip,width=8cm]{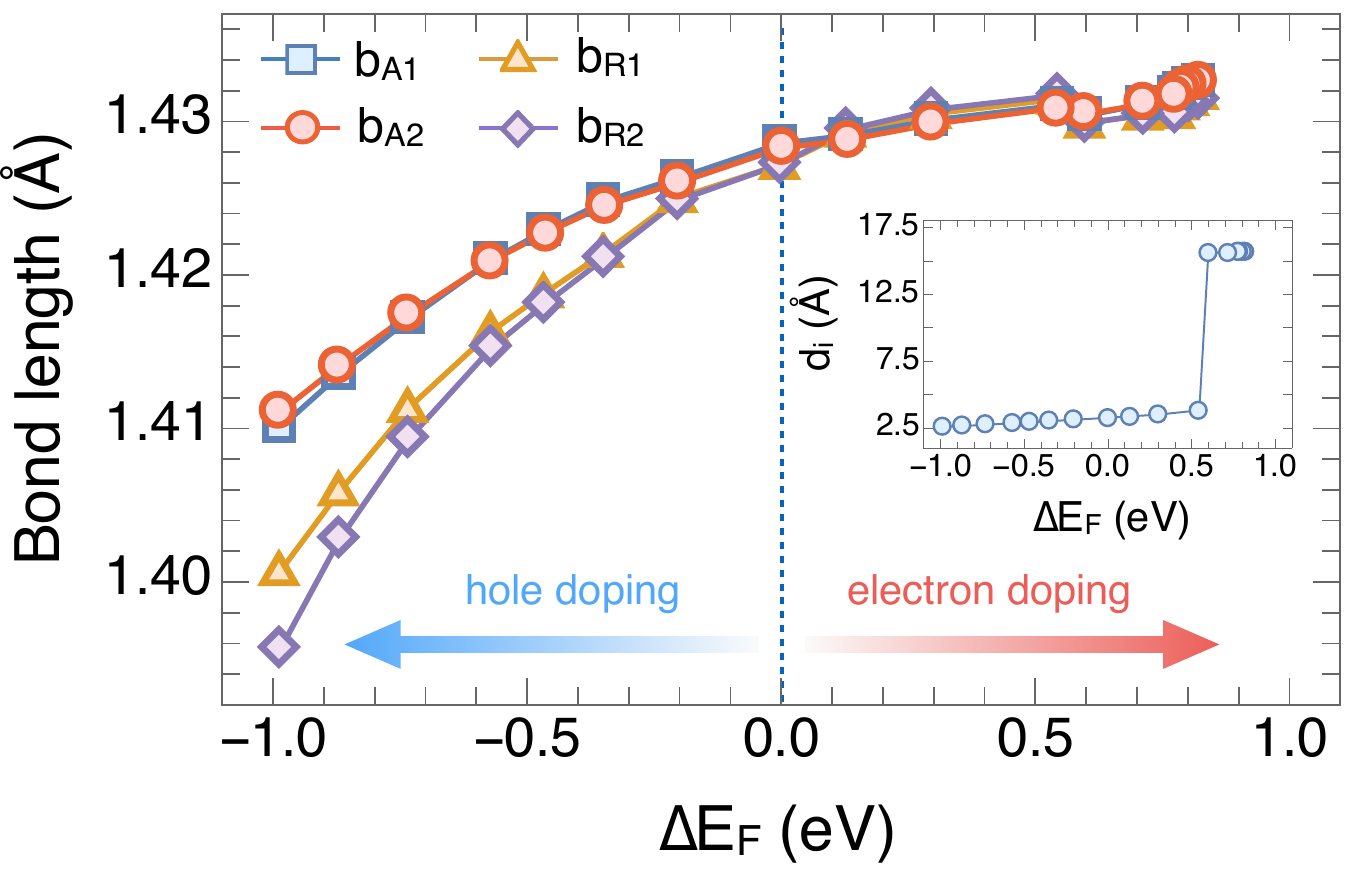}
  \caption{\label{fig:bonds} C-C bond lengths $b_{A1}$,
    $b_{A2}$, $b_{R1}$, and $b_{R2}$ as a function of
    $\Delta E_{\mathrm F}$. Insert shows the intertube distance $d_i$
    as a function of $\Delta E_{\mathrm F}$.}
\end{figure}

In Fig.~\ref{fig:bonds}, we plot the C-C bond lengths of $b_{A1}$,
$b_{A2}$, $b_{R1}$, and $b_{R2}$ [defined in Fig.~\ref{fig:model}(b)]
as a function of $\Delta E_{\mathrm F}$. The inset shows the intertube
distance $d_i$ as a function of $\Delta E_{\mathrm F}$.  The $(6, 6)$
armchair SWNT bundle has the $C_{6h}$ symmetry, thereby we consider
all different bond length variables ($b_{A1}$, $b_{A2}$, $b_{R1}$ and
$b_{R2}$) in one sixth of the bundle (totally 24 C-C bond lengths).
In the neutral case at $\Delta E_{\mathrm F}=0$, we have
$b_{A1}=b_{A2}=1.428$ \AA\ and $b_{R1}=b_{R2}=1.427$ \AA\, which are
larger than the bond lengths of the isolated $(6, 6)$ armchair SWNT
($b_{A1}=b_{A2}=1.425$ \AA\ and $b_{R1}=b_{R2}=1.424$ \AA).  It has
been reported that in the isolated armchair tubes, the bonds parallel
to the tube axis $b_{A}$ are longer than those perpendicular to the
tube axis $b_{R}$ due to the curvature effect of the
tubes~\cite{hung2016intrinsic,akai2005electronic}.  For the electron
doping, $b_{A1}$, $b_{A2}$, $b_{R1}$ and $b_{R2}$ increase by
increasing $\Delta E_{\mathrm F}$ up to 0.60 eV.  The difference
between bond lengths are not significant for
$\Delta E_{\mathrm F} > 0$.  As we can see from the inset of
Fig.~\ref{fig:bonds}, since the intertube distance suddenly increases
at $\Delta E_{\mathrm F} = 0.60$ eV under the heavy electron doping,
making a failure of the structure of the bundle with the absence of
the van der Waals interaction.  In fact, the structure of the
electron-doped $(6, 6)$ armchair SWNT bundle for
$\Delta E_{\mathrm F} \geq 0.60$ eV has higher symmetry ($D_{6h}$)
than that of the neutral one ($C_{6h}$), which leads to decrease of
$E_g$ under the electron doping (see in Fig.~\ref{fig:band-gap}).

\begin{figure}[t]
  \centering
  \includegraphics[clip,width=8cm]{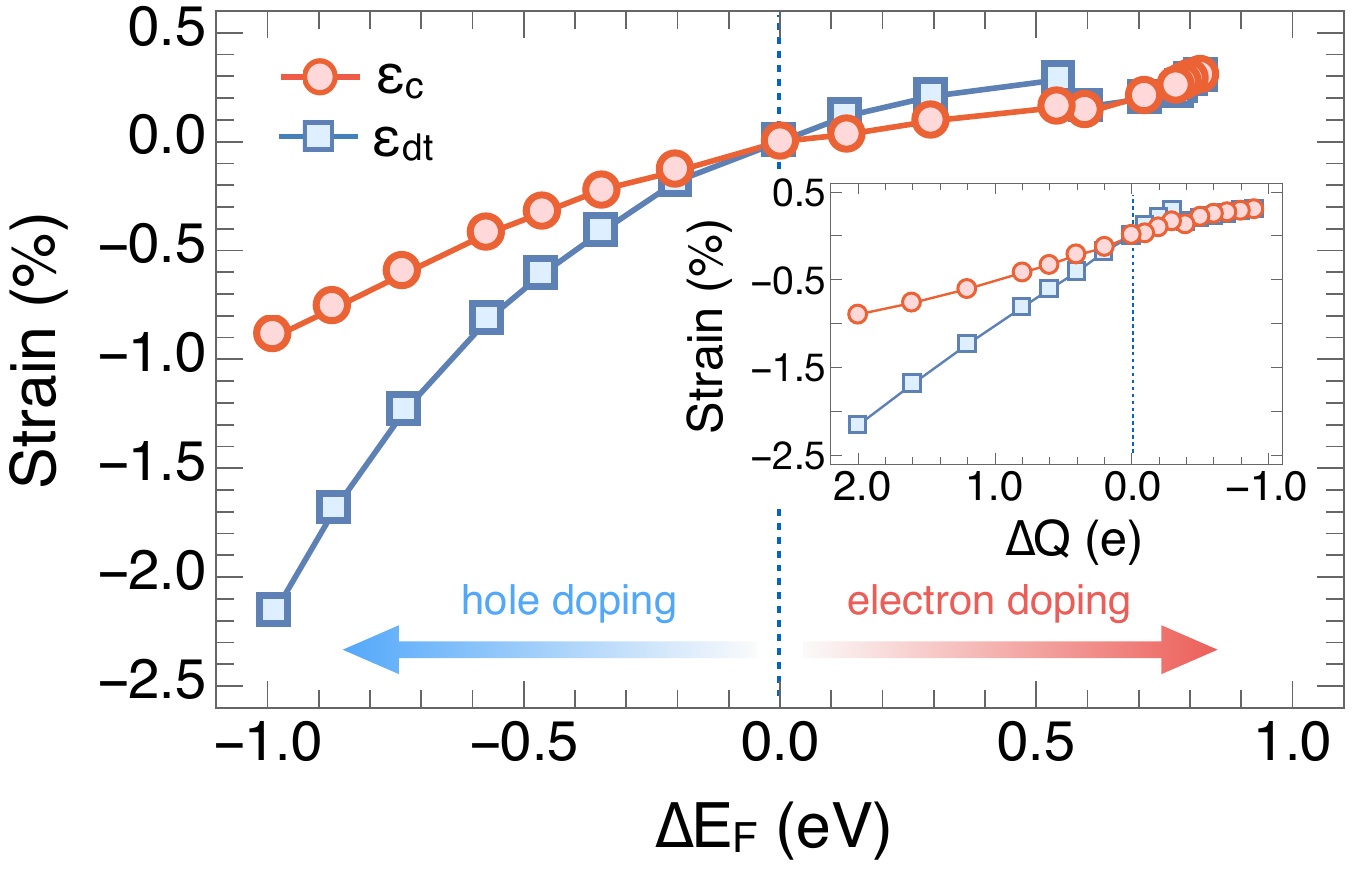}
  \caption{\label{fig:strain} Strain of length ($\varepsilon_c$) and
    strain of diameter ($\varepsilon_{dt}$) plotted as function of
    $\Delta E_{\mathrm F}$.  Inset gives the strain versus the charge
    doping.}
\end{figure}

In order to study the variation of the structural deformation as a
function of Fermi energy, we define the strain as
\begin{equation}
\varepsilon_x=\Delta x/x_0,
\end{equation}
where $x_0$ is either the length $c$ or diameter $d_t$ of the SWNT in
the unit cell of bundle at geometry optimization for neutral case, as
shown in Fig.~\ref{fig:model}(b), and $\Delta x$ is the increment of
the length $\Delta c$ or diameter $\Delta d_t$ under the charge doping
level. In Fig.~\ref{fig:strain} we plot the strain of length
$\varepsilon_c$ and diameter $\varepsilon_{dt}$ of the (6, 6) armchair
SWNT bundle as function of $\Delta E_{\mathrm F}$.  In the neutral
case, we obtain $\varepsilon_c=\varepsilon_{dt}=0$, while for the
electron (hole) doping, $\varepsilon_c$ and $\varepsilon_{dt}$
increase (decrease) with increasing (decreasing)
$\Delta E_{\mathrm F}$. In the doped SWNT bundles, the increase of
$\varepsilon_c$ and $\varepsilon_{dt}$ with the electron doping is
smaller than the decrease of $\varepsilon_c$ and $\varepsilon_{dt}$
with the hole doping. As shown in the inset of Fig.~\ref{fig:strain},
$\varepsilon_c$ and $\varepsilon_{dt}$ are approximately a linear
function of $\Delta Q$ but are nonlinear with respect to
$\Delta E_{\mathrm F}$.  At hole doping $\Delta Q = +2.0e$ (removed
$\sim 0.08$ electron per carbon atom), $\varepsilon_c\sim -1\%$ is
larger than that of individual (6, 6) SWNTs
($\varepsilon_c\sim -0.7\%$ at
$\Delta Q = +2.0e$)~\cite{sun2002dimensional}.  It is important to
note that the strain of the SWNT bundle in Fig.~\ref{fig:strain} shows
an asymmetric behavior with respect to the doping level, indicated by
a higher (smaller) strain under heavy hole (electron) doping.  On the
other hand, the strain of the individual SWNT exhibits a symmetric
shape with nearly the same strain under both electron and hole
dopings~\cite{yin2011fermi,sun2002dimensional}.  Recently, the
influence of the gate voltage on the structural deformation in both
multiwall carbon nanotube (MWNT) bundles (or yarns) and SWNT bundles
have been studied experimentally by using electrochemical
doping~\cite{foroughi2011torsional,mirfakhrai2007electrochemical,baughman1999carbon,shang2015large}.
In these experiments, the largest strain obtained is nearly $-1$\%
with the applied potential of about $-1$V (hole-doped), while the
strain is quite small with the applied potential of about $+1$V
(electron-doped)~\cite{foroughi2011torsional}.  This asymmetrical
behavior of the actuation strain is in good agreement with our present
calculation.  It should be noted that the strain range of natural
muscles is very high (up to 20\%~\cite{madden2004artificial}) compared
with that of SWNT bundles ($\sim$1\%).  Therefore, a possible approach
to obtain carbon nanotube bundle actuators is by fabricating bundle
networks from a forest of multi-walled carbon nanotubes, which could
yield strain as high as 220\% and strain rate of about
$3.7\times 10^4$\% per second~\cite{aliev2009giant}.

\begin{figure}[t]
  \centering
  \includegraphics[clip,width=8.4cm]{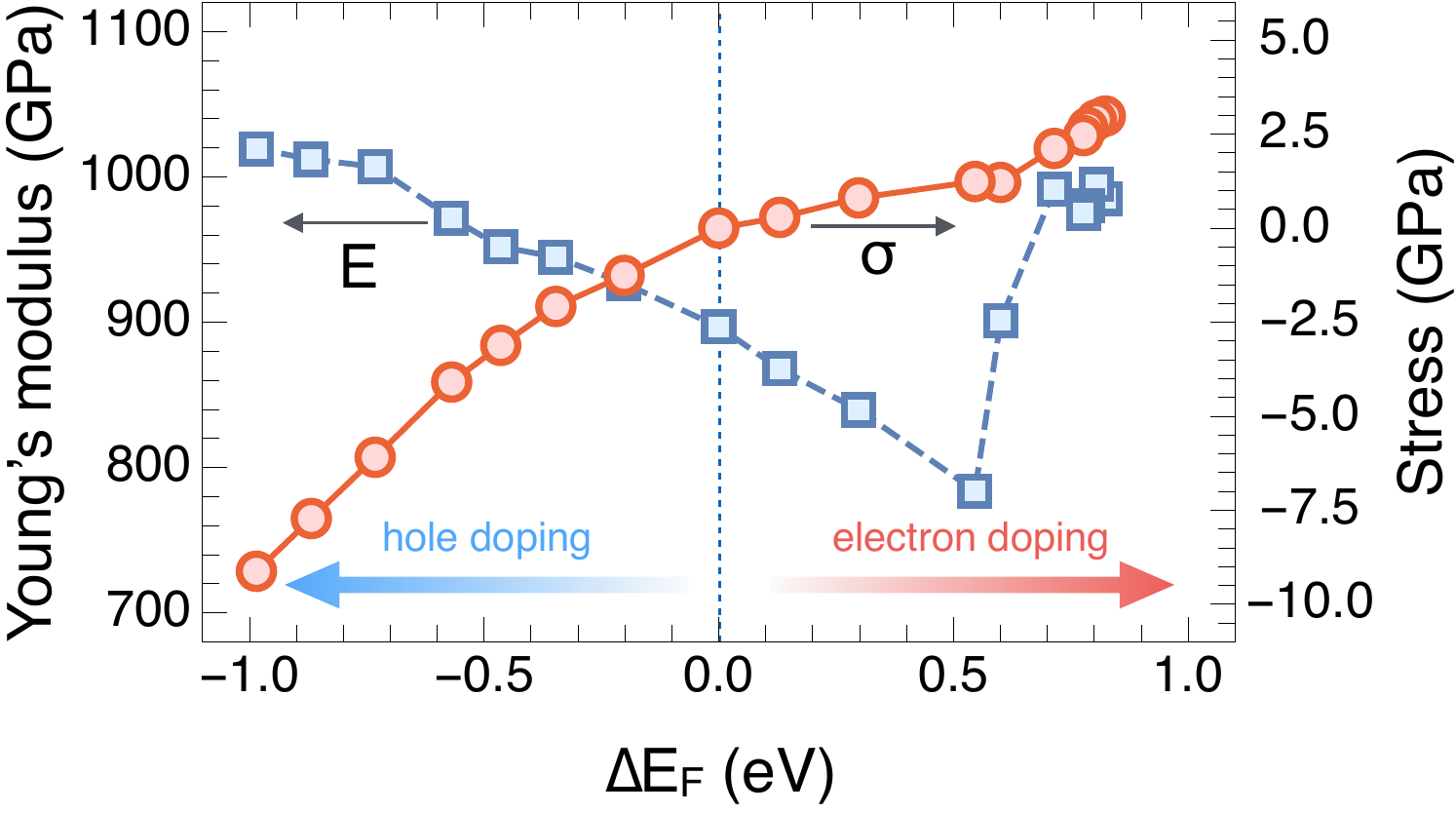}
  \caption{\label{fig:stress} Young's modulus ($E$) and stress ($\sigma$) as function of $\Delta E_{\mathrm F}$.}
\end{figure}

Finally, to study the Young modulus as a function of Fermi energy, the
SWNT bundle is initially relaxed to its minimum total energy with the
lattice vector $c$ along the axis for each $\Delta Q$ [see again
Fig.~\ref{fig:energy}(a)].  We then apply a series of small tensile
strains ($\pm 0.2\%, \pm 0.5\%, \pm 0.9\%, \pm 1.4\%, \pm 2\%$) on the
unit cell and simultaneously relax the other stress components to zero
(Poisson's ratio contraction under uniaxial tension).  Since the
system stays in the harmonic regime with the small strains, total
energy values can be fitted to a polynomial of strain.  Then, we
obtain the Young modulus $E$ from the strain derivatives of the
polynomial for each charge doping level, which is defined by
\begin{equation}
\label{eq:young}
E = \left. {\frac{1}{V_0}}
  {\frac{\partial^2U(\varepsilon)}{\partial \varepsilon^2}}
\right\vert_{\varepsilon=0} ,
\end{equation}
with
\begin{equation}
  \label{eq:strain-energy}
  U(\varepsilon) = U_0 + \frac{\partial U(\varepsilon)}{\partial
    \varepsilon}
  \varepsilon+\frac{1}{2}
  \frac{\partial U^2(\varepsilon)}{\partial
    \varepsilon^2}\varepsilon^2 
  + \dots ,
\end{equation}
where $U(\varepsilon)$, $U_0$, $V_0$, and $\varepsilon$ are the strain
energy, the total energy at equilibrium, the volume at equilibrium,
and the tensile strain for each $\Delta Q$, respectively.  To obtain
$V_0$, we use the constant thickness of a SWNT ($d_0 = 3.4$ \AA) in
the bundle~\cite{hung2016intrinsic,collins2001engineering}, in which
$d_0$ is assumed to be independent of the small strain and $\Delta Q$.
It is important to know how much force per cross-sectional area (also
known as stress) is generated by the charge doping level.  For small
strain, the axial stress can be described as $\sigma=\varepsilon_c E$,
where $\varepsilon_c$ is the axial strain as a function of
$\Delta E_{\mathrm F}$ [see in Fig.~\ref{fig:strain}] and $E$ is the
Young modulus [Eq.~\ref{eq:young}].

Figure~\ref{fig:stress} displays the Young modulus $E$ and the axial
stress $\sigma$ as a function of $\Delta E_{\mathrm F}$.  In the
neutral case, the strength of the $(6, 6)$ SWNT bundle ($E=897$ GPa)
is weaker than that of the isolated $(6, 6)$ SWNT ($E=978$
GPa~\cite{hung2016intrinsic}) because the bond lengths of the bundle
are only changed a little under the van der Waals forces between the
tubes in the bundle, as shown in Fig.~\ref{fig:bonds}. For the
electron doping case, $E$ decreases with increasing
$\Delta E_{\mathrm F}$ up to 0.60 eV. Since the structure of the
bundle is broken for heavy electron doping, the strength of the
$(6, 6)$ SWNT bundle is comparable to that of the isolated $(6, 6)$
SWNT with $\Delta E_{\mathrm F} \geq 0.60$ eV.  For the hole doping
case, $E$ increases with decreasing $\Delta E_{\mathrm F}$ because the
bundle is in a state of compression.  Large strain and large Young's
modulus lead to large stress, $\sigma \approx -9$ GPa (26000 times the
natural muscle), that is generated by heavy doping level at
$\Delta E_{\mathrm F}\approx-1$ eV.  For heavy electron doping, we
obtain $\sigma \approx 3$ GPa, weaker than in the case of hole doping.
Our results suggest that the hole doping should be good for the
artificial muscle application of the SWNT bundle.


\section{Conclusions}
We have performed a first principles theoretical study on the
structural deformation and on the electronic structure as a function
of the Fermi energy for armchair SWNT bundles.  The results obtained
reveal that the $(6, 6)$ armchair SWNT bundle is semiconducting with
energy band gap of about $0.41$ eV around the charge neutral condition
due to their low-symmetry.  The electronic properties of a $(6, 6)$
armchair SWNT bundle strongly depend on the Fermi energy.  The
Fermi-energy-dependent electronic properties give a
semiconductor-metal transition in the bundle as a result of heavy
electron doping when the relative shift of the Fermi energy is
adjusted up to 0.60 eV.  Moreover, the lengths of the C-C bonds
parallel and perpendicular to the bundle axis also depend on the Fermi
energy, resulting in the observed asymmetry of the strain-Fermi energy
curves.  For heavy hole doping, the strains of length and diameter can
be achieved up to 1\% and 2\%, respectively, while the structure of
the bundle is broken for heavy electron doping.  Because of large
strain and large Young's modulus, the stress generated by heavy hole
doping is larger than that by heavy electron doping.  This study gives
a theoretical support for the actuation response of the carbon
nanotube bundles that are tunable by the charge doping level.

\section*{Acknowledgements}
N.T.H. and A.R.T.N acknowledge the Interdepartmental Doctoral Degree
Program for Multidimensional Materials Science Leaders in Tohoku
University. R.S. acknowledges JSPS KAKENHI Grant Numbers JP25107005
and JP25286005.

\section*{References}

\bibliographystyle{elsarticle-num}

\begin{thebibliography}{100}

\bibitem{yu2000strength}
M.~F. Yu, O.~Lourie, M.~J. Dyer, K.~Moloni, T.~F. Kelly, R.~S. Ruoff, Strength
  and breaking mechanism of multiwalled carbon nanotubes under tensile load,
  Science 287 (2000) 637--640.

\bibitem{yu2000tensile}
M.~F. Yu, B.~S. Files, S.~Arepalli, R.~S. Ruoff, Tensile loading of ropes of
  single wall carbon nanotubes and their mechanical properties, Phys. Rev.
  Lett. 84 (2000) 5552.

\bibitem{hung2016intrinsic}
N.~T. Hung, D.~V. Truong, V.~V. Thanh, R.~Saito, Intrinsic strength and failure
  behaviors of ultra-small single-walled carbon nanotubes, Comput. Mater. Sci.
  114 (2016) 167--171.

\bibitem{baughman1999carbon}
R.~H. Baughman, C.~Cui, A.~A. Zakhidov, Z.~Iqbal, J.~N. Barisci, G.~M. Spinks,
  G.~G. Wallace, A.~Mazzoldi, D.~De~Rossi, A.~G. Rinzler, O.~Jaschinski,
  S.~Roth, M.~Kertesz, Carbon nanotube actuators, Science 284 (1999)
  1340--1344.

\bibitem{baughman2002carbon}
R.~H. Baughman, A.~A. Zakhidov, W.~A. de~Heer, Carbon nanotubes--the route
  toward applications, Science 297 (2002) 787--792.

\bibitem{madden2004artificial}
J.~D.~W. Madden, N.~A. Vandesteeg, P.~A. Anquetil, P.~G.~A. Madden, A.~Takshi,
  R.~Z. Pytel, S.~R. Lafontaine, P.~A. Wieringa, I.~W. Hunter, Artificial
  muscle technology: physical principles and naval prospects, IEEE J. Oceanic
  Eng. 29 (2004) 706--728.

\bibitem{gartstein2002charge}
Y.~N. Gartstein, A.~A. Zakhidov, R.~H. Baughman, Charge-induced anisotropic
  distortions of semiconducting and metallic carbon nanotubes, Phys. Rev. Lett.
  89 (2002) 045503.

\bibitem{yin2011fermi}
L.~C. Yin, H.~M. Cheng, R.~Saito, M.~S. Dresselhaus, Fermi level dependent
  optical transition energy in metallic single-walled carbon nanotubes, Carbon
  49 (2011) 4774--4780.

\bibitem{yin2010fermi}
L.~C. Yin, R.~Saito, M.~S. Dresselhaus, The fermi level dependent electronic
  properties of the smallest $(2, 2)$ carbon nanotube, Nano. Lett. 10 (2010)
  3290--3296.

\bibitem{li2007theoretical}
C.~Li, T.~W. Chou, Theoretical studies on the charge-induced failure of
  single-walled carbon nanotubes, Carbon 45 (2007) 922--930.

\bibitem{saito1992electronic}
R.~Saito, M.~Fujita, G.~Dresselhaus, M.~S. Dresselhaus, Electronic structures
  of carbon tubules based on $\textrm{C}_{60}$, Phys. Rev. B 46 (1992)
  1804--1811.

\bibitem{blum2011selective}
C.~Blum, N.~Sturzl, F.~Hennrich, S.~Lebedkin, S.~Heeg, H.~Dumlich, S.~Reich,
  M.~M. Kappes, Selective bundling of zigzag single-walled carbon nanotubes,
  ACS Nano 5 (2011) 2847--2854.

\bibitem{haroz2010enrichment}
E.~H. H{\'a}roz, W.~D. Rice, B.~Y. Lu, S.~Ghosh, R.~H. Hauge, R.~B. Weisman,
  S.~K. Doorn, J.~Kono, Enrichment of armchair carbon nanotubes via density
  gradient ultracentrifugation: Raman spectroscopy evidence, ACS Nano 4 (2010)
  1955--1962.

\bibitem{liu2011large}
H.~Liu, D.~Nishide, T.~Tanaka, H.~Kataura, Large-scale single-chirality
  separation of single-wall carbon nanotubes by simple gel chromatography, Nat.
  Commun. 2 (2011) 309.

\bibitem{shaver2008alignment}
J.~Shaver, A.~N.~G. Parra-Vasquez, S.~Hansel, O.~Portugall, C.~H. Mielke,
  M.~Von~Ortenberg, R.~H. Hauge, M.~Pasquali, J.~Kono, Alignment dynamics of
  single-walled carbon nanotubes in pulsed ultrahigh magnetic fields, ACS Nano
  3 (2008) 131--138.

\bibitem{delaney1998broken}
P.~Delaney, H.~J. Choi, J.~Ihm, S.~G. Louie, M.~L. Cohen, Broken symmetry and
  pseudogaps in ropes of carbon nanotubes, Nature 391 (1998) 466--468.

\bibitem{lankhorst96-rigidband}
M.~H.~R. Lankhorst, H.~J.~M. Bouwmeester, H.~Verweij, Use of the rigid band
  formalism to interpret the relationship between o chemical potential and
  electron concentration in $\textrm{La}_{1-x} \textrm{Sr}_x
  \mathrm{CoO}_{3-\delta}$, Phys. Rev. Lett. 77 (1996) 2989--2992.

\bibitem{marshak84-rigidband}
A.~H. Marshak, C.~M.~V. Vliet, Electrical current and carrier density in
  degenerate materials with nonuniform band structure, Proceedings of the IEEE
  72 (1984) 148--164.

\bibitem{mirfakhrai2008electromechanical}
T.~Mirfakhrai, R.~Krishna-Prasad, A.~Nojeh, J.~D.~W. Madden, Electromechanical
  actuation of single-walled carbon nanotubes: an ab initio study,
  Nanotechnology 19 (2008) 315706.

\bibitem{sun2002dimensional}
G.~Sun, J.~K{\"u}rti, M.~Kertesz, R.~H. Baughman, Dimensional changes as a
  function of charge injection in single-walled carbon nanotubes, J. Am. Chem.
  Soc. 124 (2002) 15076--15080.

\bibitem{mirfakhrai2007electrochemical}
T.~Mirfakhrai, J.~Oh, M.~Kozlov, E.~C.~W. Fok, M.~Zhang, S.~Fang, R.~H.
  Baughman, J.~D.~W. Madden, Electrochemical actuation of carbon nanotube
  yarns, Smart Mater. Struct. 16 (2007) S243.

\bibitem{foroughi2011torsional}
J.~Foroughi, G.~M. Spinks, G.~G. Wallace, J.~Oh, M.~E. Kozlov, S.~Fang,
  T.~Mirfakhrai, J.~D.~W. Madden, M.~K. Shin, S.~J. Kim, R.~H. Baughman,
  Torsional carbon nanotube artificial muscles, Science 334 (2011) 494--497.

\bibitem{giannozzi2009quantum}
P.~Giannozzi, S.~Baroni, N.~Bonini, M.~Calandra, R.~Car, C.~Cavazzoni,
  D.~Ceresoli, G.~L. Chiarotti, M.~Cococcioni, I.~Dabo, A.~D. Corso, S.~d.
  Gironcoli, S.~Fabris, G.~Fratesi, R.~Gebauer, U.~Gerstmann, C.~Gougoussis,
  A.~Kokalj, M.~Lazzeri, L.~Martin-Samos, N.~Marzari, F.~Mauri, R.~Mazzarello,
  S.~Paolini, A.~Pasquarello, L.~Paulatto, C.~Sbraccia, S.~Scandolo,
  G.~Sclauzero, A.~P. Seitsonen, A.~Smogunov, P.~Umari, R.~M. Wentzcovitch,
  Quantum espresso: a modular and open-source software project for quantum
  simulations of materials, J. Phys. Condens. Matter 21 (2009) 395502.

\bibitem{hohenberg1964inhomogeneous}
P.~Hohenberg, W.~Kohn, Inhomogeneous electron gas, Phys. Rev. 136 (1964) B864.

\bibitem{kohn1965self}
W.~Kohn, L.~J. Sham, Self-consistent equations including exchange and
  correlation effects, Phys. Rev. 140 (1965) A1133.

\bibitem{rappe1990optimized}
A.~M. Rappe, K.~M. Rabe, E.~Kaxiras, J.~D. Joannopoulos, Optimized
  pseudopotentials, Phys. Rev. B 41 (1990) 1227.

\bibitem{pseudo}
We used the pseudopotentials from http://www.quantum-espresso.org.

\bibitem{perdew1996generalized}
J.~P. Perdew, K.~Burke, M.~Ernzerhof, Generalized gradient approximation made
  simple, Phys. Rev. Lett. 77 (1996) 3865.

\bibitem{thonhauser2007van}
T.~Thonhauser, V.~R. Cooper, S.~Li, A.~Puzder, P.~Hyldgaard, D.~C. Langreth,
  Van der waals density functional: Self-consistent potential and the nature of
  the van der waals bond, Phys. Rev. B 76 (2007) 125112.

\bibitem{dumlich2011nanotube}
H.~Dumlich, S.~Reich, Nanotube bundles and tube-tube orientation: A van der
  waals density functional study, Phys. Rev. B 84 (2011) 064121.

\bibitem{monkhorst1976special}
H.~J. Monkhorst, J.~D. Pack, Special points for brillouin-zone integrations,
  Phys. Rev. B 13 (1976) 5188.

\bibitem{broyden1970convergence}
C.~G. Broyden, The convergence of a class of double-rank minimization
  algorithms 1. general considerations, IMA J. Appl. Math. 6 (1970) 76--90.

\bibitem{fletcher1970new}
R.~Fletcher, A new approach to variable metric algorithms, Comput. J. 13 (1970)
  317--322.

\bibitem{goldfarb1970family}
D.~Goldfarb, A family of variable-metric methods derived by variational means,
  Math. Comput. 24 (1970) 23--26.

\bibitem{shanno1970conditioning}
D.~F. Shanno, Conditioning of quasi-newton methods for function minimization,
  Math. Comput. 24 (1970) 647--656.

\bibitem{okada2001pressure}
S.~Okada, A.~Oshiyama, S.~Saito, Pressure and orientation effects on the
  electronic structure of carbon nanotube bundles, J. Phys. Soc. Jpn. 70 (2001)
  2345--2352.

\bibitem{saito1998physical}
R.~Saito, G.~Dresselhaus, M.~S. Dresselhaus, Physical Properties of Carbon
  Nanotubes, World Scientific, 1998.

\bibitem{reich2002electronic}
S.~Reich, C.~Thomsen, P.~Ordej{\'o}n, Electronic band structure of isolated and
  bundled carbon nanotubes, Phys. Rev. B 65 (2002) 155411.

\bibitem{ajayan1999nanotubes}
P.~M. Ajayan, Nanotubes from carbon, Chem. Rev. 99 (1999) 1787.

\bibitem{okada2000nearly}
S.~Okada, A.~Oshiyama, S.~Susumu, Nearly free electron states in 
  carbon nanotube bundles, Phys. Rev. B 62 (2000) 7634.

\bibitem{ayala2010doping}
P.~Ajayan, R.~Arenal, M.~R{\"u}mmeli, A.~Rubio, P.~Pichler, The 
  doping of carbon nanotubes with nitrogen and their potential 
  applications, Carbon 99 (2010) 575--586.

\bibitem{hung2015diameter}
N.~T. Hung, A.~R. T. Nugraha, E.~H. Hasdeo, M.~S. Dresselhaus, R.~Saito, 
  Diameter dependence of thermoelectric power of semiconducting 
  carbon nanotubes, Phys. Rev. B 92 (2015) 165426.

\bibitem{akai2005electronic}
Y.~Akai, S.~Saito, Electronic structure, energetics and geometric structure of
  carbon nanotubes: A density-functional study, Physica E 29 (2005) 555--559.

\bibitem{shang2015large}
Y.~Shang, X.~He, C.~Wang, L.~Zhu, Q.~Peng, E.~Shi, S.~Wu, Y.~Yang, W.~Xu,
  R.~Wang, S.~Du, A.~Cao, Y.~Li, Large-deformation, multifunctional artificial
  muscles based on single-walled carbon nanotube yarns, Adv. Eng. Mater. 17
  (2015) 14--20.

\bibitem{aliev2009giant}
A.~E. Aliev, J.~Oh, M.~E. Kozlov, A.~A. Kuznetsov, S.~Fang,
  A.~F. Fonseca, R.~Ovalle, M.~D. Lima, M.~H. Haque, Y.~N. Gartstein,
  M.~Zhang, A.~A. Zakhidov, R.~H. Baughman, Giant-stroke, superelastic
  carbon nanotube aerogel muscles, Science 323 (2009) 1575--1578.

\bibitem{collins2001engineering}
P.~G. Collins, M.~S. Arnold, P.~Avouris, Engineering carbon nanotubes and
  nanotube circuits using electrical breakdown, Science 292 (2001) 706--709.

\end{thebibliography}

\end{document}